\documentclass[5p]{elsarticle}

\usepackage{setspace}
\usepackage{color}
\usepackage{epsf}
\usepackage{graphicx}

\newcommand{\revis}[1]{\textcolor{black}{#1}}
\begin{document}
\title{Advanced multicanonical Monte Carlo methods for efficient simulations of nucleation processes of polymers}
\author[1]{Stefan Schnabel\corref{cor1}}
\ead{stefanschnabel@physast.uga.edu}
\author[2]{Wolfhard Janke}
\ead{Wolfhard.Janke@itp.uni-leipzig.de}
\author[3]{Michael Bachmann}
\ead{m.bachmann@fz-juelich.de}
\cortext[cor1]{Corresponding author}

\address[1]{Center for Simulational Physics, University of Georgia, Athens, GA 30602, USA}

\address[2]{Institut f\"ur Theoretische Physik and Centre for Theoretical Sciences (NTZ),\\
Universit\"at Leipzig, Postfach 100920, D-04009 Leipzig, Germany}

\address[3]{Soft Matter Systems Research Group, Institut f\"ur Festk\"orperforschung (IFF-2) and Institute for Advanced Simulation (IAS-2), 
Forschungszentrum J\"ulich, D-52425 J\"ulich, Germany}

%\ead[url]{http://www.physik.uni-leipzig.de/CQT.html}

%
%
\begin{abstract}
The investigation of freezing transitions of single polymers is computationally demanding, since surface effects dominate the nucleation process. In recent studies we have systematically shown that the freezing properties of flexible, elastic polymers depend on the precise chain length. Performing multicanonical Monte Carlo simulations, we faced several computational challenges in connection with liquid-solid and solid-solid transitions. For this reason, we developed novel methods and update strategies to overcome the arising problems.
%Here, we explain the details of our simulation techniques.
\revis{We introduce novel Monte Carlo moves and two extensions to the multicanonical method.}
\end{abstract}
\maketitle
\section{Introduction}
Induced by the rapidly increasing efficiency and availability of computational resources, the field of computational physics has gained tremendously in importance within the last decades, and it is today regarded as physics' third pillar alongside experimental and theoretical physics. In addition to the innovations in hardware, simulation techniques have evolved further, and in fact, the greater improvements have resulted from better methods rather than from faster computers. A particularly important application is the investigation of thermodynamic properties of complex systems by means of Markov chain Monte Carlo methods. Starting sixty years ago with the Metropolis algorithm \cite{metropolis}, which emulates the canonical ensemble, the arsenal of algorithms has been extended and more sophisticated methods have been introduced. Among the most powerful simulation techniques are generalized-ensemble\linebreak methods such as parallel tempering \cite{parallel_temp1,parallel_temp2}, multicanonical sampling \cite{muca}, simulated tempering \cite{simulated_temp}, or the Wang-Landau method \cite{wanglandau}, which allow in principle to collect all information about the entire thermodynamic behavior of the investigated system in a single simulation. However, depending on the considered system, substantial difficulties can occur, part of which are specifically related to properties of the system being studied, whereas others, like broken ergodicity, are of more general nature.

In a recent study on flexible homopolymers \cite{FENE_cpl,FENE_jcp}, we encountered a number of problems of both kinds and developed new simulation techniques to overcome these. Some of them are rather specific to polymers, while others are more general and can also be applied to nonmolecular systems. Combining our strategies we were able to boost the efficiency of our algorithms and to perform very precise simulations of systems which could not be investigated in this quality before.

The purpose of this paper is to explain our methods in detail. After a short introduction of the applied polymer model in the next section, we briefly explain in section 3 the multicanonical Monte Carlo method, which served as the basic algorithm in our simulations. The following section 4 is dedicated to the applied conformational updates and includes a new general optimization strategy for basic updates of systems with continuous degrees of freedom. Afterwards we introduce and motivate in section 5 two general extensions to the multicanonical method, and finish in section 6 with some concluding remarks.

\section{Model}
In our simulations we employed a bead-spring model for flexible, elastic polymers. For a specified set of monomer coordinates $\{\mathbf{X}\}$, the energy of a polymer conformation is given by
\begin{eqnarray}
E(\{\mathbf{X}\})=\sum\limits_{i=1}^{N-1}\sum\limits_{j=i+1}^{N}E_{\rm nb}(|\mathbf{X}_i-\mathbf{X}_j|)\nonumber\\
+\sum\limits_{i=1}^{N-1}E_{\rm b}(|\mathbf{X}_{i+1}-\mathbf{X}_{i}|).
\end{eqnarray}
Here, the non-bonded interaction
\begin{equation}
 		E_{\rm nb}(r)=E_{\rm LJ}(\min\{r,r_{\rm c}\})-E_{\rm LJ}(r_{\rm c})
\end{equation}
corresponds to a truncated and shifted Lennard-Jones (LJ) potential
\begin{equation}
 		E_{\rm LJ}(r)=4[(\sigma/r)^{12}-(\sigma/r)^{6}] 
\end{equation}
with the cutoff radius $r_{\rm c}$. Pairs of bonded monomers further interact via
\begin{equation}
 		E_{\rm b}(r)=-\frac{K}{2}R^2\ln(1-[(r-r_0)/R]^2),
\end{equation}
which is the standard finitely extensible non-linear elastic (FENE) potential. The parameters are chosen such that the minima of both potentials coincide at $r_0$, in order to prevent frustration. For details of the parametrization see \cite{FENE_jcp,binder}.

\revis{This model belongs to the class of coarse-grained models, i.e., microscopic details have been traded for generality and handiness. However, accurate simulations are still a substantial challenge.}

\section{Multicanonical Monte Carlo Sampling}
Before we discuss our novel simulation strategies, let us first recall basic principles of Markov chain Monte Carlo simulations \revis{\cite{Hastings}}, for which acceptance criteria are obtained from the master equation:
\begin{equation}
\frac{dP_\mu (t)}{dt}=\sum\limits_\nu [P_\nu (t)W_{\nu \rightarrow \mu}-P_\mu (t)W_{\mu \rightarrow \nu}],
\end{equation}
where $P_\mu (t)$ denotes the probability for a state $\mu$ to occur at time $t$ and $W_{\nu\rightarrow\mu}$ is the transition probability from state $\nu$ to $\mu$. In stationary equilibrium, where $dP_\mu (t)/dt=0$, this equation is solved by:
\begin{equation}
P_\nu W_{\nu \rightarrow \mu}=P_\mu W_{\mu \rightarrow \nu},
\label{eqn:det_bal}
\end{equation}
called  ``detailed balance''. The transition probability $W_{\nu \rightarrow \mu}$ is the product of the probability of {\em selecting\/} the update proposal $W^{\rm s}_{\nu \rightarrow \mu}$ and the probability $W^{\rm a}_{\nu \rightarrow \mu}$ of {\em accepting\/} it:
\begin{equation}
W_{\nu \rightarrow \mu}=W^{\rm s}_{\nu \rightarrow \mu}W^{\rm a}_{\nu \rightarrow \mu}.
\end{equation}
Symmetric selection probabilities
\begin{equation}
W^{\rm s}_{\nu \rightarrow \mu}=W^{\rm s}_{\mu \rightarrow \nu},
\label{eqn:sym_Wprop}
\end{equation}
entail
\begin{equation}
\frac{W^{\rm a}_{\nu \rightarrow \mu}}{W^{\rm a}_{\mu \rightarrow \nu}}=\frac{P_\mu}{P_\nu},
\label{eqn:W_a_ratio}
\end{equation}
for which the most common solution is given by
\begin{equation}
W^{\rm a}_{\nu \rightarrow \mu}=\min\left(1,\frac{P_\mu}{P_\nu}\right).
\label{eqn:Wa_sym}
\end{equation}
\revis{However, for convenience or increased sampling efficiency, it is useful to introduce Monte Carlo updates where the selection probabilities are unequal:
\begin{equation}
W^{\rm s}_{\nu \rightarrow \mu}\ne W^{\rm s}_{\mu \rightarrow \nu},
\end{equation}
in which case
\begin{equation}
\frac{W^{\rm a}_{\nu \rightarrow \mu}}{W^{\rm a}_{\mu \rightarrow \nu}}=\frac{P_\mu W^{\rm s}_{\mu \rightarrow \nu}}{P_\nu W^{\rm s}_{\nu \rightarrow \mu}}.
\label{eqn:Wa_ratio_asym}
\end{equation}
Then, the more general expression
\begin{equation}
W^{\rm a}_{\nu \rightarrow \mu}=\min\left(1,\frac{P_\mu W^{\rm s}_{\mu \rightarrow \nu}}{P_\nu W^{\rm s}_{\nu \rightarrow \mu}}\right).
\label{eqn:Wa_asym}
\end{equation}
of the acceptance probability is required. It has been demonstrated that such weighted updates can enable a much more efficient sampling of the system conformations \cite{up_irbaeck}, compared with symmetrically chosen selection probabilities. This also applies to simulations in the grand-canonical ensemble (constant chemical potential) or a constant pressure in the $N\!pt$ ensemble, where volume fluctuations are relevant \cite{Mezei2}.}

The goal of the multicanonical method \cite{muca} is to generate a flat histogram $H$ over a certain macroscopic observable which in our case is the energy $E$. This is achieved by introducing a weight function $\omega(E)$ which is inversely proportional to the density of states $g(E)$:
\begin{eqnarray}
H(E)&=&\mathrm{const}=\omega(E)g(E),\\
\omega(E)&\propto&g^{-1}(E).
\end{eqnarray}
A single point in state space (conformation) $\mu=\{\mathbf{X}\}$ is in the multicanonical ensemble represented by a probability density which is proportional to the weight function and is therefore depending only on the energy:
\begin{equation}
P_{\{\mathbf{X}\}}\propto \omega(E(\{\mathbf{X}\})).
\label{eqn:P_muca}
\end{equation}
The acceptance probability for a proposed Monte Carlo move is according to (\ref{eqn:Wa_asym}) 
\begin{equation}
\kern-5mm
W^{\rm a}_{\{\mathbf{X}\}\rightarrow\{\mathbf{X}'\}}=\min\left(1,\frac{\omega(E(\{\mathbf{X}'\}))W^{\rm s}_{\{\mathbf{X}'\}\rightarrow\{\mathbf{X}\}}}{\omega(E(\{\mathbf{X}\}))W^{\rm s}_{\{\mathbf{X}\}\rightarrow\{\mathbf{X}'\}}}\right).
\end{equation}
Usually, the density of states and hence the weight function is not known in the beginning and has to be estimated by iterative procedures such as error weighted accumulation \cite{muca_it} or the Wang-Landau method \cite{wanglandau}.

\section{Conformational Update Proposals}
\subsection{Displacement move with energy dependent maximal step length}
When investigating many-particle systems by means of Monte Carlo simulations, the simplest possible conformational update is the displacement of a single particle to a uniformly distributed random position $\mathbf{X}_i'$ within a sphere\footnote{Instead of a sphere, any three-dimensional body which is invariant under inversion of coordinates, e.g., an adequately oriented cube, would serve as well.} around its original location $\mathbf{X}_i$:
\begin{equation}
	\mathbf{X}_i'=\mathbf{X}_i+\mathbf{r}, {\rm \ with\ }	|\mathbf{r}|\le r_{\rm max}.
\end{equation}
In the case of a flexible polymer with elastic bonds, such updates can be applied to all monomers. Thereby, the size $r_{\rm max}$ of the sphere crucially influences the performance of the simulation. A larger sphere allows the system to perform extended steps in conformational space and is therefore appropriate for simulations at high temperatures. If the temperature is lowered, the efficiency decreases since the proposed steps are now too large, and the system will not smoothly descend to narrow local energy minima. Moreover, if the system eventually finds an energy minimum, further moves are unlikely to be accepted, since the proposed changes will almost certainly result in a huge increase in energy. In consequence, smaller spheres should be used when a system with a rough energy landscape is investigated at low temperatures. It is simple to incorporate variable sphere radii into simulation techniques such as Metropolis \cite{metropolis}, parallel tempering \cite{parallel_temp2}, or simulated tempering \cite{simulated_temp} by assigning suitable sphere radii to each temperature, i.e., to use $r_{\rm max}(T)$ instead of $r_{\rm max}$, since for each of these methods a (sub)ensemble is associated to each single temperature and detailed balance is satisfied. Changes in temperature are usually performed separately from moves in conformational space and hence need not to be considered here.

The situation is more complicated for multicanonical and Wang-Landau sampling, where a simulation temperature does not exist. Instead, the entire state space is sampled in a single generalized ensemble, making it difficult to choose a single sphere radius that leads to adequate performance. However, the application of variable sphere radii is highly desirable, as it would greatly improve simulation efficiency. Since we require large steps at high and small steps at low  energies, the energy itself appears to be a well-suited control parameter for the sphere radii. However, using the standard multicanonical method with a maximal step length that depends on energy, and therefore changes in time, would cause a violation of the detailed balance condition. 

Let us discuss this in more detail by considering a displacement of the $k$th monomer. Assume a conformation $\{\mathbf{X}^h\}$ with a certain relatively high energy $E_h$, and assume further, the maximum step length $r_{\rm max}(E_h)$, is comparatively large. During the following update the system might jump to a rather small energy $E_l$ with a much smaller sphere radius $r_{\rm max}(E_l)<r_{\rm max}(E_h)$. That means the maximum step length for the next update is smaller than for the first. As one consequence, the system sometimes cannot reach the starting point $\mathbf{X}_k^h$ within a single step, hence detailed balance is clearly violated. This is the case if the distance between the two positions exceeds the smaller sphere radius  $|\mathbf{X}_k^h-\mathbf{X}_k^l|>r_{\rm max}(E_l)$. Note that $|\mathbf{X}_k^h-\mathbf{X}_k^l|\le r_{\rm max}(E_h)$ holds by definition. Even if this is not the case and the starting point lies within the smaller sphere, detailed balance is not fulfilled, because the probability densities for selecting the forward and the backward update are different and (\ref{eqn:sym_Wprop}) is violated. Fortunately, according to (\ref{eqn:Wa_asym}), the emerging bias can easily be corrected. The probability density of proposing a certain displacement equals the inverse volume of the sphere:
{\small
\begin{equation}
\kern-5mm
W^{\rm s}_{\nu \rightarrow \mu}=\left\{ \begin{array}{l}
1/(\frac{4}{3}\pi r_{\rm max}^3(E_\nu)), {\rm if}\ |\mathbf{X}_k^\nu-\mathbf{X}_k^\mu|\le r_{\rm max}(E_\nu)\\
0,\ {\rm else}.
\end{array} \right.
\end{equation}}
\noindent For $|\mathbf{X}_k^\nu-\mathbf{X}_k^\mu|\le r_{\rm max}(E_\nu)$, one obtains according to (\ref{eqn:Wa_ratio_asym})
\begin{equation}
\frac{W^{\rm a}_{\nu \rightarrow \mu}}{W^{\rm a}_{\mu \rightarrow \nu}}=\frac{P_\mu r^{-3}_{\rm max}(E_\mu)}{P_\nu r^{-3}_{\rm max}(E_\nu)}.
\end{equation}
Hence, the final acceptance criterion reads
{\small
\begin{equation}
\label{accept_prop_shiftUp}
\kern-8.5mm
W^{\rm a}_{\nu \rightarrow \mu} \! = \! \left\{\!\! \begin{array}{l}
\!\min\left(1,\frac{P_\mu r_{\rm max}^3(E_\nu)}{P_\nu r_{\rm max}^3(E_\mu)}\right)\!, {\rm if}\ |\mathbf{X}_k^\nu-\mathbf{X}_k^\mu|\le r_{\rm max}(E_\mu)\\
\!0,\ {\rm else}.
\end{array} \right .
\end{equation}}
\noindent Remember that the case $|\mathbf{X}_k^\nu-\mathbf{X}_k^\mu|> r_{\rm max}(E_\nu)$ cannot occur and is therefore not considered.

\begin{figure}

\begin{center}
\begin{minipage}{0.45\textwidth}
{
\includegraphics[width=\textwidth]{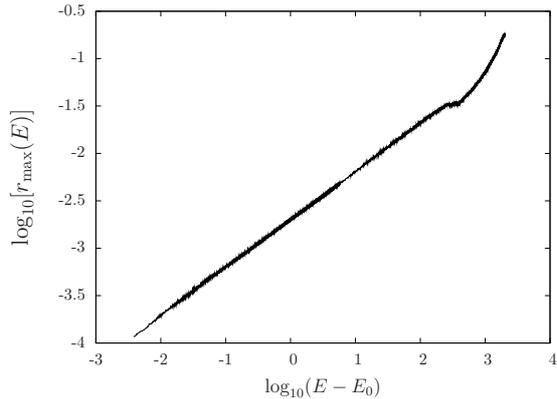}
}
\end{minipage}
\caption{\small{\label{fig:r_max} \emph{Maximal step length $r_{\rm max}(E)$ after a preliminary tuning procedure for  $N=309\ (E_0=-1820.684)$.}}}
\end{center}
\end{figure}

In principle, any strictly positive function $r_{\rm max}(E)$ can be employed, but here we are searching for a function that results in appropriate acceptance rates for all energies. Therefore we start with a flat function and perform a tuning procedure. First, we apply a standard binning, i.e., we divide the energy axis in intervals in which $r_{\rm max}(E)$ is constant, i.e., if $E_i\le E< E_{i+1}$ then $r_{\rm max}(E)=r_{\rm max}(E_i)$, with a fixed interval size $\Delta E=E_{i+1}-E_i$. The value of $r_{\rm max}(E_i)$ shall now be adjusted such that approximately two third of all proposed updates increase the energy while the remaining third leads to lower energies. It is reasonable to assume that such values for $r_{\rm max}(E_i)$ exist, since for very small values the accessible part of the energy landscape resembles a tilted hyperplane with one half belonging to higher and the other half to lower energies. If on the other hand $r_{\rm max}(E_i)$ is very large, the great majority of accessible states will have higher energies, because the density of states usually decreases rapidly with energy. In consequence, there must be a value of $r_{\rm max}(E_i)$ in-between that shows the desired property. In order to find this value we modify the radii after any proposed update 
%$\{\mathbf{X}\}\rightarrow\{\mathbf{X'}\}$
$\nu\rightarrow\mu$
according to
%{\small
\begin{equation}
\label{r_max_tune}
%\kern-8mm
r'_{\rm max}(E_i)=\left\{ \begin{array}{l}
(1- \epsilon)r_{\rm max}(E_i), {\rm if}\ E_\nu\le E_\mu\\
(1+2\epsilon)r_{\rm max}(E_i), {\rm if}\ E_\nu>E_\mu,
%(1- \epsilon)r_{\rm max}(E_i), {\rm if}\ E(\{\mathbf{X}\})\le E(\{\mathbf{X'}\})\\
%(1+2\epsilon)r_{\rm max}(E_i), {\rm if}\ E(\{\mathbf{X}\})>E(\{\mathbf{X'}\}),
\end{array} \right .
\end{equation}
%}
\noindent with $E_i<E_\nu<E_{i+1}$ and $0<\epsilon\ll 1$.
%\noindent with $E_i<E(\{\mathbf{X}\})<E_{i+1}$ and $0<\epsilon\ll 1$.
 It is easy to see that $r_{\rm max}(E_i)$ will remain approximately unaltered if it has the desired characteristics, i.e., if $E_\nu<E_\mu$ in 66.6\% of all cases. If the fraction of proposed moves leading to higher energies is too big, $r_{\rm max}$ will be reduced and if it is too small, $r_{\rm max}$ will be increased. In our simulation we used $\epsilon=10^{-3}\ldots 10^{-5}$ and found little difference in performance. As expected, higher values of $\epsilon$ allow faster convergence but lead to more noise in $r_{\rm max}(E_i)$. However, in all considered cases $r_{\rm max}(E_i)$ converged quickly and caused update acceptance rates above 60\% for all energies. In Fig.~\ref{fig:r_max}, the obtained radii for the homopolymer of length $N=309$ are shown. The used ratio 1:2 was chosen for the sake of simplicity. Different values might be found to be appropriate as well. The only restriction is that the desired fraction of updates to higher energies must be larger than $1/2$.

\begin{figure}
\begin{center}
\begin{minipage}{0.45\textwidth}
{
\includegraphics[width=\textwidth]{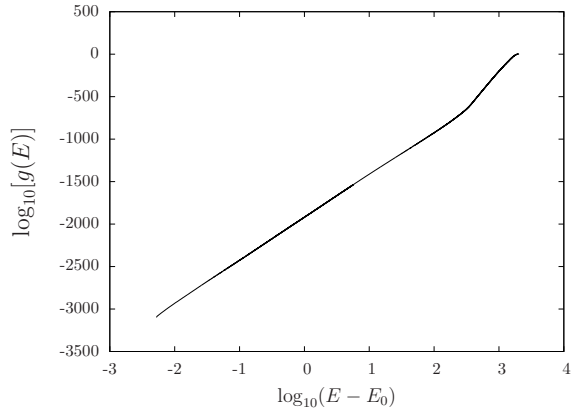}
}
\end{minipage}
\caption{\small{\label{fig:zstd309} \emph{Density of states for $N=309\ (E_0=-1820.684)$, covering more then 3000 orders of magnitude.}}}
\end{center}
\end{figure}

If the applied algorithm is able to find the valley of the global energy minimum, in principle the optimization allows us to come arbitrarily close to the ground state. Remaining problems are of ``technical'' nature and consider the resolution of the energy scale and limits of numerical data types. In Fig.~\ref{fig:zstd309}, the density of states $g(E)$ for the 309mer as obtained from two simulations is shown. After we investigated the general behavior and covered approximately 2000 orders of magnitude in the density of states, we resampled the region $E<-1815$ with a much higher energy resolution gaining further 1000 orders of magnitude in $g(E)$.

%The energy dependent sphere radius can also serve as an easily obtained rough estimate for the vibrational freedom of the particles and thus according to the Lindemann criterion reveals which energies are likely to belong to solid states.

\revis{In a similar approach, attempted some time ago \cite{Mezei}, the authors applied analytic functions $r_{\rm max}(e_k)$ depending on the energy $e_k$ of the single particle $k$ that is to be moved within a canonical ensemble. In contrast to the results presented here, decisive improvements could not be achieved. Most likely this is in the first place due to the fact that in the canonical ensemble the potential for speedups is much smaller than in the multicanonical ensemble. We also believe for two reasons that the energy of a single particle as the argument of $r_{\rm max}$ is in general less favorable then the energy of the entire system. First, when the system approaches the ground state, the particles might possess differing energies but $r_{\rm max}$ has to be close to zero for all of them. Secondly, the same displacement will cause smaller relative changes for the global energy than for the single-particle energy and, therefore, smaller changes in $r_{\rm max}$. Thus, the correction factor will be closer to unity if the global energy is used and the general acceptance will be higher and/or larger steps are possible.}

Notice that the described tuning procedure leads to a violation of the detailed balance condition which seemed to be of little relevance, though, presumably since $\epsilon$ is small. Of course, the tuning must be ceased for the production run, in order to exclude this source of systematic error.

If the considered system has continuous degrees of freedom, this optimization procedure should in principle always be applicable to basic Monte Carlo moves. However, the method might not work as described in the exceptional situations when the density of states {\em decreases\/} with increasing energy. In these (rare) cases one should not rely on the proposed energy, but on the density of states itself, i.e., the radius has to be reduced (increased) if the update leads to an energy with a higher (lower) density of states. This was not necessary for the here investigated polymer model and since the density of states is not known a priori we employed the energy as reference.

\subsection{Bond-exchange moves}
While performing bond-exchange moves the positions of the monomers remain unchanged, but the bonds between them are rearranged. In the past this type of conformational update has been applied mainly to lattice polymers \cite{cutjoin_hp}, and 
%recently,
applications for off-lattice polymers have also been documented and proven to be efficient \revis{\cite{Theodorou,cutjoin_contin}}. For the sake of completeness we present the two different types used in our investigations.

\begin{figure}
\begin{center}
\begin{minipage}{0.45\textwidth}
{
\includegraphics[width=\textwidth]{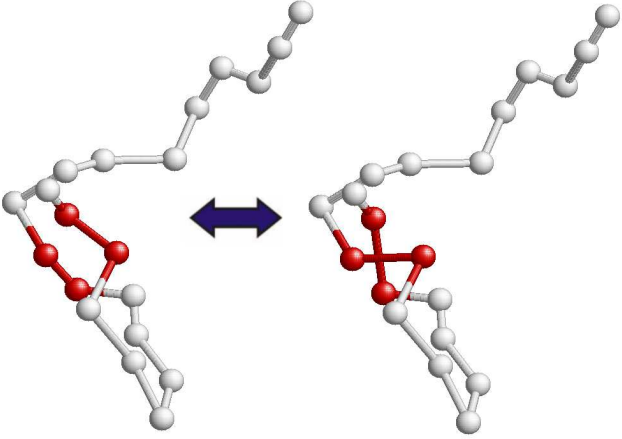}
}
\end{minipage}
\caption{\small{\label{fig:up_bond} \emph{Bond-exchange update.}}}
\end{center}
\end{figure}

The first version, depicted in Fig.~\ref{fig:up_bond}, consists of a swap of bonds between four nearby monomers. Initially, the monomers are labeled by numbers according to their position along the chain. Assuming two bonds have been chosen to be swapped, only one way exists to reconnect the chain without splitting the polymer. Let the contributing monomers be on the positions $i,i+1,j$, and $j+1$ with $j>i+1$. It is obvious that if the $i$th bond between monomer $i$ and $i+1$ and the $j$th bond between monomer $j$ and $j+1$ are removed, different bonds can only be established between the $i$th and the $j$th monomer on the one side, and between the $(i+1)$th and the $(j+1)$th monomer on the other. Creating a bond between the $(i+1)$th and the $j$th monomer would result in a closed loop, since both monomers are already connected by a sequence of bonds. In our simulations, we first randomly choose an arbitrary bond $i$ and determine afterwards which other bonds can possibly participate in an exchange update. Since in the employed model the bond length has an upper and a lower limit, only a few bonds are candidates. From this group the second bond $j$ is then drawn randomly and the acceptance probability is calculated.

At this point it is important to recognize that also for this type of Monte Carlo move the probability for selecting the update, which is inversely proportional to the product of the number of bonds $N_{\rm bonds}$ and the number of possible exchange partner bonds $n^{\rm epb}_{\nu,i}$, often differs from that of the backward update. Both must be calculated and used for the determination of the acceptance probability according to (\ref{eqn:Wa_asym}). One obtains
\begin{eqnarray}
W^{\rm a}_{\nu \rightarrow \mu}&=&\min\left(1,\frac{P_\mu W^{\rm s}_{\mu \rightarrow \nu}}{P_\nu W^{\rm s}_{\nu \rightarrow \mu}}\right),\nonumber\\
	&=&\min\left(1,\frac{P_\mu\cdot (N_{\rm bonds}n^{\rm epb}_{\nu,i})^{-1}}{P_\nu\cdot (N_{\rm bonds}n^{\rm epb}_{\mu,i})^{-1}}\right),\nonumber\\
	&=&\min\left(1,\frac{P_\mu n^{\rm epb}_{\mu,i}}{P_\nu n^{\rm epb}_{\nu,i}}\right).
\label{accept_prop_bondexcUp}
\end{eqnarray}
The order of monomers and bonds gets changed during the update and eventually appears to be totally random, if it is not restored by relabeling.

\begin{figure}
\begin{center}
\begin{minipage}{0.45\textwidth}
{
\includegraphics[width=\textwidth]{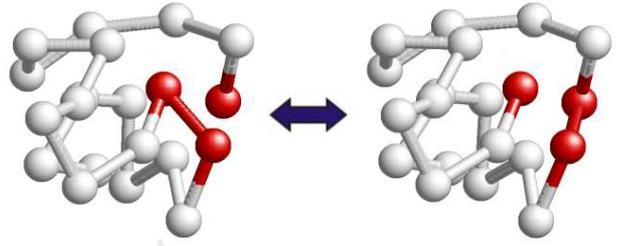}
}
\end{minipage}
\caption{\small{\label{fig:up_bond_end} \emph{End-bond-exchange update.}}}
\end{center}
\end{figure}

If only the update just described is used, an end mono\-mer will always remain an end monomer and the simulation would still be inefficient. Hence, we applied a second bond-exchange move shown in Fig.~\ref{fig:up_bond_end}. Thereby we connect an end monomer to another nearby monomer and break the created loop by removing the old bond next to the formed junction. More explicitly, if we connect the first monomer to the $j$th, we obtain a ring of bonds connecting the first $j$ monomers with a side chain branching off at the $j$th monomer. To remove the junction we have to delete the bond between the $(j-1)$th and the $j$th monomer. In the second case where the $N$th monomer gets connected to the $j$th, the bond between the monomers $j$ and $j+1$ has to be deleted. Within the simulation we choose one of the end monomers and determine all monomers that are possible partners for the update. Again, we draw monomer $j$ from this set and, in order to be able to calculate the acceptance probability $W^a$, it is necessary  to consider the selection probabilities for the update in both directions:
\begin{equation}
W^{\rm a}_{\nu \rightarrow \mu}=\min\left(1,\frac{P_\mu n^{\rm epm}_{\mu,i}}{P_\nu n^{\rm epm}_{\nu,i}}\right),
\label{accept_prop_endexcUp}
\end{equation}
with $n^{\rm epm}_{i,\nu}$ and $n^{\rm epm}_{i,\mu}$ being the numbers of possible exchange partner monomers and $i\in\{1,N\}$.

\begin{figure}
\begin{center}
\begin{minipage}{0.45\textwidth}
{
\includegraphics[width=\textwidth]{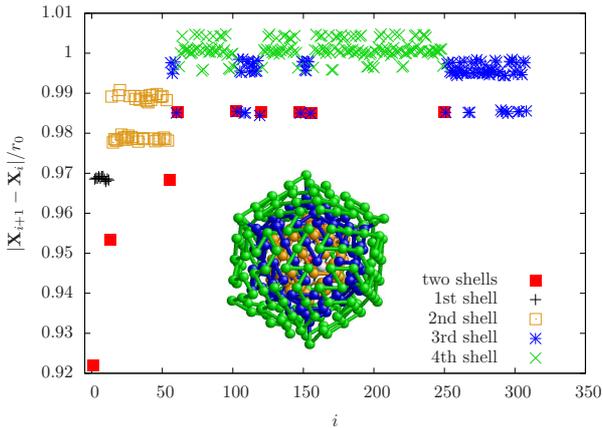}
}
\end{minipage}
\caption{\small{\label{fig:bond_length_309} \emph{Bond ordering for the putative ground state of the 309mer icosahedron.}}}
\end{center}
\end{figure}

The application of the two bond-exchange updates significantly increased the performance of the simulation and allowed larger changes of the polymer's configuration also in the ``frozen'' low-temperature regime. Even if there are no noticeable changes in monomer positions, the bonds are still quite flexible and arrange in a specific order when zero temperature is approached. Exemplified for the lowest-energy conformation of the 309mer, the length of each bond is shown in Fig.~\ref{fig:bond_length_309}, where the shell to which it belongs is represented by the symbol and the color. As a result of the icosahedral packing, neighboring monomers are closest when they belong to neighboring shells. This makes these monomer pairs unfavorable for bonds, and in consequence only one bond each connects the inner shells, and at low $T$ one end of the polymer is always located in the center.

\subsection{Monomer cut-and-paste update}
\begin{figure}
\begin{center}
\begin{minipage}{0.45\textwidth}
\begin{center}
{
\includegraphics[width=\textwidth]{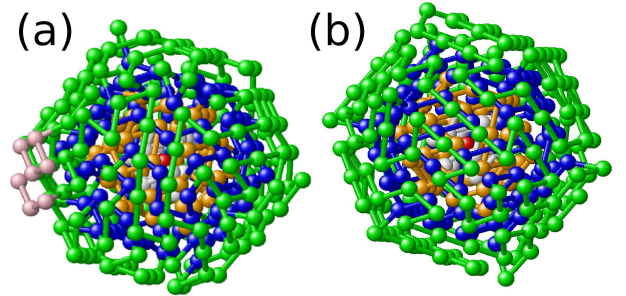}
}
\end{center}
\end{minipage}
\caption{\small{\label{fig:309_defects} \emph{(a) Polymer with length $N=309$ at low temperature forming an icosahedron with a surface defect, (b) ground state conformation. For the color code of the shells, cp. Fig.~5.}}}
\end{center}
\end{figure}

\begin{figure}
\begin{center}
\begin{minipage}{0.45\textwidth}
{
\includegraphics[width=\textwidth]{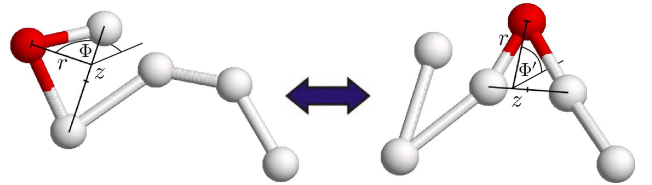}
}
\end{minipage}
\caption{\small{\label{fig:up_be} \emph{Monomer cut-and-paste update.}}}
\end{center}
\end{figure}

Below the liquid-solid transition the representative conformations differ not only in the arrangement of the bonds, but in monomer positions as well. Even if the ground state is a perfect icosahedron, single monomers can be displaced at low temperatures, thereby creating multiple surface defects (Fig.~\ref{fig:309_defects}). Transitions between these microstates cannot be performed with simple monomer displacements and bond-exchange moves only, since high energy barriers separate favorable monomer positions on the surface of the icosahedron. Hence, we developed a fourth type of Monte Carlo move (Fig.~\ref{fig:up_be}) to overcome this difficulty. For this update, a monomer $i$ is selected whose neighbors are at a appropriate distance to be bonded themselves. In order to possess two neighbors the chosen monomer must not be an end of the polymer ($1\neq i\neq N$). The position $(r,\phi,z)_s$ of monomer $i$ is then determined according to a cylindrical coordinate system $s$ defined as follows: The $z$-axis points through the neighboring monomers $i-1$ and $i+1$ and the origin is located in their midpoint. The further orientation of $s$ is irrelevant, because the original angle $\phi$ will not be needed in the following. Now, monomer $i$ is cut and a bond connecting the monomers $i-1$ and $i+1$ is created while another existing bond is removed in order to paste monomer $i$ at its position. For that purpose, a second coordinate system $s'$ is defined similar to $s$ but based on the adjacent monomers of the removed bond, say monomer $j$ and $j+1$ ($j\neq i\neq j+1$). The coordinates $r$ and $z$ are now transposed from $s$ to $s'$ and a new angle $\phi'$ is drawn randomly from $[0,2\pi)$. Again the angular orientation of $s'$ can be arbitrary. Monomer $i$ is now placed at this new position $(r,\phi',z)_{s'}$ and connected to the monomers $j$ and $j+1$.

The selection probabilities of the move and its inversion are identical, and no correction needs to be applied at this point. However, it is appropriate to introduce restrictions to the choice of the monomer to be moved and the bond to be split. If the polymer occupies a compact shape, the update has only a good chance of acceptance when performed at the surface, since moving a monomer within the interior, as well as from the center to the surface, implies a large increase in energy and a very low acceptance rate. It is therefore useful to choose only bonds and monomers that are in regions of minor density, e.g., at the surface of a compact conformation. To estimate the density we use the number of contacts of a monomer (for details see \cite{FENE_cpl}), i.e., the number of monomers to which its distance does not exceed a certain threshold. Since inner monomers at low temperature always have 12 contacts, we choose only monomers with less than 11 contacts and bonds that connect monomers with less than 12 neighbors. Unfortunately this leads to unequal selection probabilities and requires once more the introduction of a correction term. If $n^{\nu}_m$ is the number of monomers to choose from and $n^{\nu}_{b,i}$ is the number of available bonds, we obtain
\begin{equation}
W^{\rm a}_{\nu \rightarrow \mu}=\min\left(1,\frac{P_\mu n^{\nu}_m n^{\nu}_{b,i}}{P_\nu n^{\mu}_m n^{\mu}_{b,i}}\right).
\end{equation}
Here, $n^{\nu}_{b,i}$ depends on $i$ in a non-trivial way since bonds adjacent to monomer $i$ must not be chosen. An alternative way would be to allow choosing these bonds, but to immediately reject the update, once they are selected.

\section{Extensions to the Multicanonical Sampling Algorithm}
In the previous section we described how to overcome the problem of energy barriers through avoiding them by the application of certain update procedures, which is possible in the described cases since the configurations on both ``sides'' of the barriers are rather similar. For the bond-exchange update the monomer positions are identical, and in the case of the cut-and-paste update, only a single monomer is moved. However, other barriers of different nature exist, and need to be treated with other strategies. As we have shown \cite{FENE_jcp}, the polymers adopt different geometries corresponding to different optimizing strategies, resembling the behavior of atomic LJ clusters. This similarity has been already reported for a slightly different model \cite{Calvo} some time ago and is the result of the matching minimum distances of the two interaction potentials, which ensure that configurations minimizing the Lennard-Jones potential also lead to low bond energies.  Clusters and polymers both favor icosahedral crystal-like conformations at temperatures below the liquid-solid-transition. These conformations divide into two subgroups according to the type of the outer layer which can be either Mackay (fcc) or anti-Mackay (hcp) \cite{northby}. Transitions between these two types occur at different temperatures, and for certain system sizes, the investigation with standard Monte Carlo methods is difficult or impossible due to high free-energy barriers between different solid phases associated with Mackay or anti-Mackay growth. A second type of solid-solid transition that occurs for special system sizes involves non-icosahedral ground-state conformations, which can be of fcc-, decahedral, or tetrahedral structure. These systems change to an icosahedral solid state at very low temperatures, posing a considerable challenge to the applied simulation method.

\subsection{``Grand-multicanonical'' simulation}
First, we will consider the Mackay--anti-Mackay transition within the surface of an icosahedral conformation. As already mentioned, the investigated LJ homopolymer behaves very similar to atomic LJ clusters. In the interval $N\in[13,147]$, we find anti-Mackay ground states for $13<N<31$ and $55<N<81$ while for the remaining polymer lengths Mackay ground states are favored. Exceptions are $N=38,75-77,86,87$  \cite{FENE_jcp}.

%\begin{figure}
%\begin{center}
%\begin{minipage}{0.45\textwidth}
%{
%\includegraphics[width=\textwidth]{./state_space2.eps}
%}
%\end{minipage}
%\caption{\small{\label{fig:state_space} \emph{Sketch of the conformational state space at low temperatures.}}}
%\end{center}
%\end{figure}

\begin{figure}
\begin{center}
\begin{minipage}{0.45\textwidth}
{
\includegraphics[width=\textwidth]{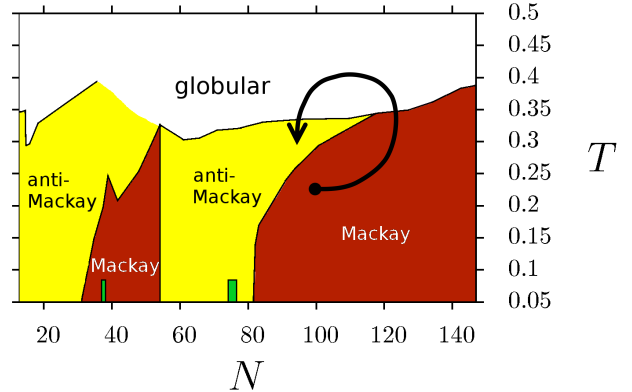}
}
\end{minipage}
\caption{\small{\label{fig:state_space_circumvent} \emph{Sketch of the conformational state space at low temperatures with ``paths'' to avoid the Mackay--anti-Mackay barrier.}}}
\end{center}
\end{figure}

Most of the systems with Mackay ground states undergo a transition to anti-Mackay conformations at a transition temperature which generally increases with system size (Fig.~\ref{fig:state_space_circumvent}). It turned out that this transition complicates the investigation, if it takes place at low temperatures, as for $N=31$, or if the system is large, e.g., for $N\ge81$. If standard methods like parallel tempering \cite{parallel_temp2}, multicanonical sampling \cite{muca} or the Wang-Landau method \cite{wanglandau} are applied, the system has to cross the barrier between the Mackay and the anti-Mackay state many times in order to produce precise results. It turned out that this can be avoided by allowing the system to move also in $N$-direction, i.e., to change its size, during the simulation. The system is then able to circumvent the Mackay--anti-Mackay transition by changing $N$, and performing two liquid-solid transitions (Fig.~\ref{fig:state_space_circumvent}), which happens more frequently than the crossing of the Mackay--anti-Mackay transition line for sizes $81\le N\le 110$.

\begin{figure}
\begin{center}
\begin{minipage}{0.45\textwidth}
{
\includegraphics[width=\textwidth]{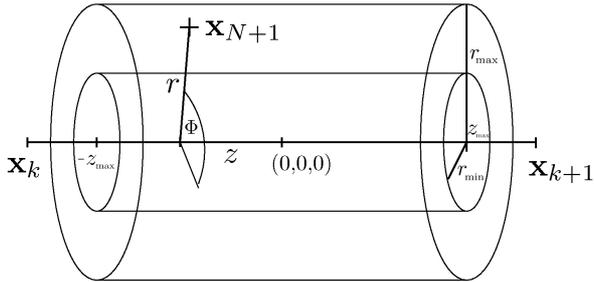}
}
\end{minipage}
\caption{\small{\label{fig:coord_up_size} \emph{Coordinates for adding a new monomer.}}}
\end{center}
\end{figure}

To move in $N$-direction we need a new Monte Carlo update that changes the system size at runtime. Fortunately, the monomer cut-and-paste update introduced above can be used as a starting point. If an increase of system size should be proposed, a bond $k$ can be picked and coordinates of the new monomer are randomized. We again apply a cylindrical coordinate system $s$, defined by the adjacent monomers of the chosen bond: the $z$-axis points through these monomers and their midpoint defines the origin. The angular orientation is arbitrary. The coordinates $(r,\phi,z)_s$ have to be determined in order to be uniformly distributed in the hollow cylinder defined by $r_{\rm min}, r_{\rm max}$, and $z_{\rm max}$ (Fig.~\ref{fig:coord_up_size}). Therefore, $\phi$ and $z$ are drawn from constant distributions over the intervals $[0,2\pi)$ and $(-z_{\rm max},z_{\rm max})$, respectively. Within $[r_{\rm min},r_{\rm max})$ the desired probability density $P_{\rm R}(r)$ has to be proportional to the area of the cylinder shell with radius $r$, i.e., proportional to $r$ itself. 
If we regard the radius $r$ as a monotonic function of a uniformly distributed random number $\xi$:
\begin{equation}
r=r(\xi),
\end{equation}
where the probability density of $\xi$ is given by
\begin{equation}
P_{\Xi}(\xi)=\left\{ \begin{array}{l}
  1,\ {\rm if}\ \in \left[0,1\right),\\
  0,\ {\rm else},
\end{array} \right .
\end{equation}
the fraction of points in the ring between $r_{\rm min}$ and $r$
\begin{equation}
f(r)=\frac{\pi(r^2-r^2_{\rm min})}{\pi(r^2_{\rm max}-r^2_{\rm min})}
\end{equation}
has to equal $\xi$,
\begin{equation}
\xi=f(r),
\end{equation}
which leads to
%we can use
%\begin{equation}
%P_{\rm R}(r)= \left(P_{\rm \Xi}(\xi)(\xi)F'^{-1}(\xi)\right)|_{\xi=F^{-1}(r)}
%\end{equation}
%to obtain
%\begin{equation}
%F(\xi)=k_1\sqrt{\xi+k_2},
%\end{equation}
%with constants $k_1$ and $k_2$. Therefore, we can derive $r$ from random numbers $\eta$, which are uniformly distributed in $[0,1)$, by
\begin{equation}
r=\sqrt{(r^2_{\rm max}-r^2_{\rm min})\xi+r^2_{\rm min}}.
\end{equation}

The inverse update meaning the reduction of the system size is simpler to accomplish. A monomer, which must not be an end monomer, is chosen randomly and once more the coordinates $(r',\phi',z')_{s'}$ in a cylindrical system $s'$ defined by the neighbors are determined. The update may only be performed if $|z'|<z_{\rm max}$ and $r_{\rm min}\le r'<r_{\rm max}$, since otherwise the inverse update would be impossible, violating detailed balance. Note that in its present form the update contains another imbalance, since for the first choice the number of alternatives differs. If the system size should be increased, we choose from $N-1$ bonds while, if the size is to be decreased, there are only $N'-2$ monomers (with $N'=N+1$) to choose from. However, this imbalance can be neglected, since it does not effect the balance of conformations with identical $N$.

To calculate the acceptance probability we first need the probability of each conformation. Again, we use a weight function $\omega(E,N)$ to produce a flat distribution but now in the two directions $N$ and $E$. It is
\begin{equation}
P_{\{\mathbf{X}\}}\propto \omega(E(\{\mathbf{X}\}),N(\{\mathbf{X}\}))
\end{equation}
and with (\ref{eqn:Wa_sym}) we easily obtain
{\small
\begin{equation}
\kern-8mm
W^{\rm a}_{\{\mathbf{X}\}\rightarrow\{\mathbf{X}'\}}=\min\left(1,\frac{\omega(E(\{\mathbf{X}'\}),N(\mathbf{X}'))W^{\rm s}_{\{\mathbf{X}'\}\rightarrow\{\mathbf{X}\}}}{\omega(E(\{\mathbf{X}\}),N(\mathbf{X}))W^{\rm s}_{\{\mathbf{X}\}\rightarrow\{\mathbf{X}'\}}}\right).
\end{equation}}
\noindent Again, it is appropriate to choose only bonds and mono\-mers from the surface. The adaptation of the method and the determination of $W^{\rm s}$ are very similar to the procedure we discussed for the cut-and-paste update and are not repeated here. Note that the imbalance  mentioned in the last paragraph is cured this way, too.

\begin{figure}[t]
\begin{center}
\begin{minipage}{0.45\textwidth}
{
\includegraphics[width=\textwidth]{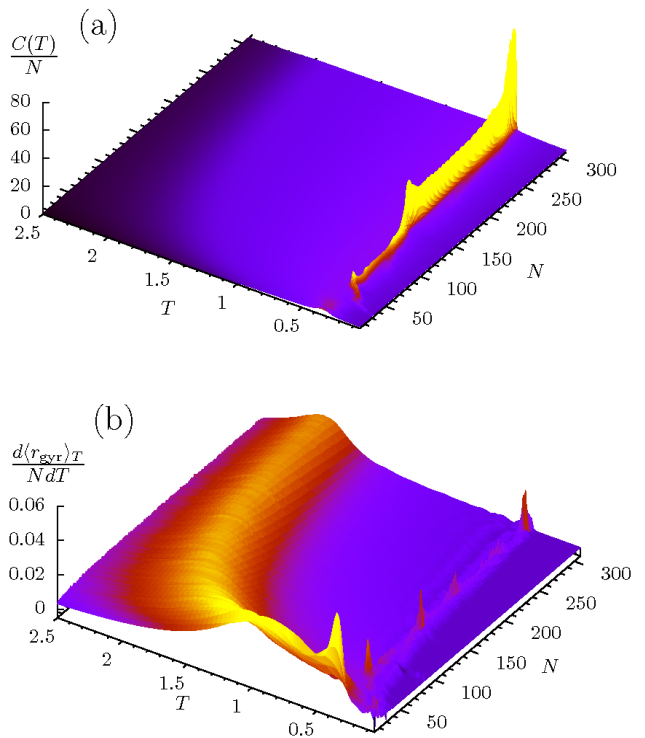}
}
\end{minipage}
\caption{\small{\label{fig:fluct} \emph{Results from a single grand-multicanonical simulation: (a) specific heat, (b) temperature derivative of the normalized radius of gyration.}}}
\end{center}
\end{figure}

This algorithm proved to be surprisingly efficient. While it appeared to be impossible to investigate the full behavior of the 100mer with standard multicanonical simulations, the simultaneous sampling of all chains with $N\le147$ did not pose any major difficulties. Furthermore, we were able to derive the thermodynamics for all polymers of size $13\le N\le309$ down to $T\approx0.05$ within a single simulation on a single Intel Xeon core (3.06GHz). This simulation involved $2\times 10^{12}$ single updates and ran for approximately 5 months. Some results are shown in Fig.~\ref{fig:fluct}.

\revis{Note that this method is primarily not designed to investigate the grand-canonical ensemble. Here, the focus is still on systems of fixed size and the merit lies in greater efficiency in sampling them simultaneously and not in a physical understanding of polymerization processes.}

\subsection{Multicanonical simulation with multiple weight functions}
The existence of non-icosahedral ground states for a\-to\-mic LJ clusters of certain sizes has been known for a long time, but the identification of these ground states is still regarded to be a major challenge to the applied algorithm. On the other hand, the investigation of the associated solid-solid transitions is even more complicated, since the goal is not only to reach the ground-state conformation but also to maintain detailed balance and to measure the density of states very precisely. To the best of our knowledge there has been only one successful attempt to solve the problem for the 98-atom cluster \cite{sharapov}, which involved the construction of an artificial energy landscape based on the prior knowledge of low-energy conformations. Here, \revis{we present} an extension to the multicanonical approach which allows for investigating the solid-solid transitions of LJ polymers and clusters, but at the same time is general enough to be of use in other cases, too.

In \cite{FENE_cpl,FENE_jcp}, we used the number of icosahedral cells to  introduce a parameter $\nu$ that indicates the geometrical state of the system: With high reliability we found $\nu=0$ for unstructured and for non-icosahedral states, $\nu=1$ for icosahedral states with Mackay overlayer, and $\nu=2$ for icosahedral states with anti-Mackay overlayer.

\begin{figure}
\begin{center}
\begin{minipage}{0.45\textwidth}
{
\includegraphics[width=\textwidth]{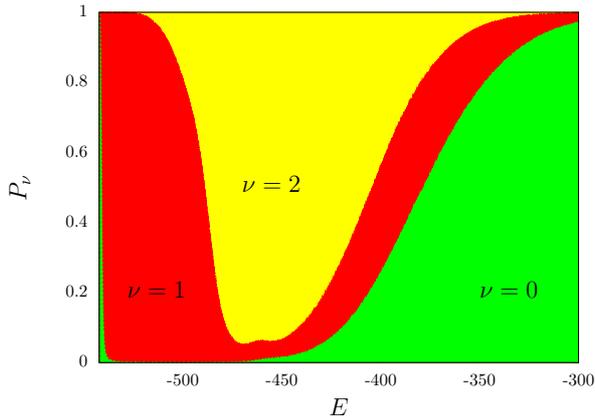}
}
\end{minipage}
\caption{\small{\label{fig:micro_ensemble} \emph{Decomposition of the microcanonical ensembles according to the different values of the order parameter $\nu$.}}}
\end{center}
\end{figure}

For the 98mer with large cutoff ($r_{\rm c}=5\sigma$) the decompositions of the ``microcanonical'' ensembles according to this parameter are depicted in Fig.~\ref{fig:micro_ensemble}. Since the different values of $\nu$ belong to very different structures, lines between the different domains in Fig.~\ref{fig:micro_ensemble} can only be penetrated in the high-energy regime. Hence, any algorithm producing these microcanonical distributions (e.g., simulated tempering, parallel tempering, the multicanonical method or the Wang-Landau technique) is prevented from finding the tetrahedral ground-state conformation, since the probability to pass through the bottle neck belonging to $\nu=0$ at $E\approx-500$ is by far too small. The solution is to balance the probabilities of the three subensembles by introducing single weight functions for each value of $\nu$. Based on the multicanonical approach (\ref{eqn:P_muca}), we use
\begin{equation}
P_{\{\mathbf{X}\}}\propto \omega_{\nu(\{\mathbf{X}\})}(E(\{\mathbf{X}\}))
\end{equation}
to derive the acceptance probability
{\small
\begin{equation}
\kern-5mm W^{\rm a}_{\{\mathbf{X}\}\rightarrow\{\mathbf{X}'\}}=\min\left(1,\frac{\omega_{\nu(\mathbf{X}')}(E(\{\mathbf{X}'\}))W^{\rm s}_{\{\mathbf{X}'\}\rightarrow\{\mathbf{X}\}}}{\omega_{\nu(\mathbf{X})}(E(\{\mathbf{X}\}))W^{\rm s}_{\{\mathbf{X}\}\rightarrow\{\mathbf{X}'\}}}\right).
\end{equation}}
\noindent The remaining task is to tune the multiple weight function $\omega$ to allow each geometry to participate equally at any energy  and to enable the system to reach the energies where the solid-solid transition takes place.

Results of applications of this algorithm are reported in detail in Ref.~\cite{FENE_jcp}.

\section{Conclusions}
In this paper, we described methods used to investigate the behavior of flexible homopolymers in much more detail and at much lower temperatures than it was previously possible.

With the energy-dependent step length we introduced a novel general optimization scheme for basic Monte Carlo moves for systems with continuous degrees of freedom\linebreak which allows constantly high acceptance rates everywhere in energy space. Applying this procedure in combination with multicanonical sampling we were able to estimate the density of states over several thousands of orders of magnitudes.

We then described two bond-exchange moves and de\-monstrated that these updates allow the reordering of polymer bonds without alteration of monomer positions. Subsequently, with the monomer-jump update we introduced a novel Monte Carlo move which increased the efficiency of the simulation further in two ways. First, the update allows the tunneling of energy barriers in the solid phase and second, it performs larger changes in the unstructured globular and the random coil phase.

By enabling variations in system size at runtime we extended the multicanonical ensemble. This led to an additional gain in efficiency since the thus modified algorithm was able to circumvent certain energy barriers or to penetrate them where they are low, i.e., at their ``weak'' points. As a result we obtained information over the entire state space over a large size interval from a single simulation.

Finally, confronted with the problem of broken ergodicity and low-temperature solid-solid transitions, we developed a second extension to the standard multicanonical technique. Due to the application of additional weight functions it is possible to retain ergodicity and to reach ``hidden'' ground states by circumventing the ``blocking'' states at intermediate temperatures. Although we yet have demonstrated the potential of this methods for hompolymers only, it is a general approach and, in combination with suitable order parameters, it might lead to substantial progress in the investigation of many other systems as well.

\section*{Acknowledgements}
We are indebted to Daniel Seaton for helpful comments and careful reading of the manuscript and thank David P. Landau for discussions. This work is partly funded by the NSF under Grant No. DMR - 0810223, the DFG under Grant Nos. JA 483/24-1/2/3, the Leipzig Graduate School of Excellence ``BuildMoNa'', the German-French DFH-UFA PhD College under Grant No. CDFA-08-07 and the John von Neumann Institute for Computing (NIC) at the Forschungszentrum J\"ulich for supercomputer time grants hlz11, jiff39, and jiff43. MB thanks the German-Israeli ``Umbrella'' consortium for support under Grant Nos. SIM6 and HPC\_2.

\end{document}